\documentclass[aps,prl,preprint,tightenlines,superscriptaddress,showpacs,byrevtex]{revtex4}
\usepackage{graphicx} 
\usepackage{dcolumn}  

\graphicspath{{ps}}

\newcommand{\ks}{K^0_S}
\newcommand{\mev}{\mathrm{MeV}/c^2}

\newcommand{\dss}{D_{s1}(2536)^+}
\newcommand{\dst}{D^{*+}}
\newcommand{\dppik}{D^+\pi^- K^+}

\begin{document}

\title{Observation of \boldmath{$D_{s1}(2536)^+ \rightarrow D^+ \pi^- K^+$} and angular decomposition of \boldmath{$D_{s1}(2536)^+\rightarrow D^{*+}K^0_S$}}
\affiliation{Budker Institute of Nuclear Physics, Novosibirsk}
\affiliation{Chiba University, Chiba}
\affiliation{University of Cincinnati, Cincinnati, Ohio 45221}
\affiliation{Department of Physics, Fu Jen Catholic University, Taipei}
\affiliation{Justus-Liebig-Universit\"at Gie\ss{}en, Gie\ss{}en}
\affiliation{The Graduate University for Advanced Studies, Hayama}
\affiliation{Hanyang University, Seoul}
\affiliation{University of Hawaii, Honolulu, Hawaii 96822}
\affiliation{High Energy Accelerator Research Organization (KEK), Tsukuba}
\affiliation{Hiroshima Institute of Technology, Hiroshima}
\affiliation{Institute of High Energy Physics, Chinese Academy of Sciences, Beijing}
\affiliation{Institute of High Energy Physics, Vienna}
\affiliation{Institute of High Energy Physics, Protvino}
\affiliation{Institute for Theoretical and Experimental Physics, Moscow}
\affiliation{J. Stefan Institute, Ljubljana}
\affiliation{Kanagawa University, Yokohama}
\affiliation{Korea University, Seoul}
\affiliation{Kyungpook National University, Taegu}
\affiliation{\'Ecole Polytechnique F\'ed\'erale de Lausanne (EPFL), Lausanne}
\affiliation{University of Ljubljana, Ljubljana}
\affiliation{University of Maribor, Maribor}
\affiliation{University of Melbourne, School of Physics, Victoria 3010}
\affiliation{Nagoya University, Nagoya}
\affiliation{Nara Women's University, Nara}
\affiliation{National Central University, Chung-li}
\affiliation{National United University, Miao Li}
\affiliation{Department of Physics, National Taiwan University, Taipei}
\affiliation{H. Niewodniczanski Institute of Nuclear Physics, Krakow}
\affiliation{Nippon Dental University, Niigata}
\affiliation{Niigata University, Niigata}
\affiliation{University of Nova Gorica, Nova Gorica}
\affiliation{Osaka City University, Osaka}
\affiliation{Osaka University, Osaka}
\affiliation{Panjab University, Chandigarh}
\affiliation{Princeton University, Princeton, New Jersey 08544}
\affiliation{Saga University, Saga}
\affiliation{University of Science and Technology of China, Hefei}
\affiliation{Seoul National University, Seoul}
\affiliation{Sungkyunkwan University, Suwon}
\affiliation{University of Sydney, Sydney, New South Wales}
\affiliation{Toho University, Funabashi}
\affiliation{Tohoku Gakuin University, Tagajo}
\affiliation{Tohoku University, Sendai}
\affiliation{Department of Physics, University of Tokyo, Tokyo}
\affiliation{Tokyo Institute of Technology, Tokyo}
\affiliation{Tokyo Metropolitan University, Tokyo}
\affiliation{Tokyo University of Agriculture and Technology, Tokyo}
\affiliation{Virginia Polytechnic Institute and State University, Blacksburg, Virginia 24061}
\affiliation{Yonsei University, Seoul}
  \author{V.~Balagura}\affiliation{Institute for Theoretical and Experimental Physics, Moscow} 
  \author{I.~Adachi}\affiliation{High Energy Accelerator Research Organization (KEK), Tsukuba} 
  \author{H.~Aihara}\affiliation{Department of Physics, University of Tokyo, Tokyo} 
  \author{K.~Arinstein}\affiliation{Budker Institute of Nuclear Physics, Novosibirsk} 
  \author{V.~Aulchenko}\affiliation{Budker Institute of Nuclear Physics, Novosibirsk} 
  \author{T.~Aushev}\affiliation{\'Ecole Polytechnique F\'ed\'erale de Lausanne (EPFL), Lausanne}\affiliation{Institute for Theoretical and Experimental Physics, Moscow} 
  \author{A.~M.~Bakich}\affiliation{University of Sydney, Sydney, New South Wales} 
  \author{E.~Barberio}\affiliation{University of Melbourne, School of Physics, Victoria 3010} 
  \author{A.~Bay}\affiliation{\'Ecole Polytechnique F\'ed\'erale de Lausanne (EPFL), Lausanne} 
  \author{K.~Belous}\affiliation{Institute of High Energy Physics, Protvino} 
% ...
\collaboration{The Belle Collaboration}

\begin{abstract}
Using 462~fb$^{-1}$ of $e^+e^-$ annihilation data recorded by the Belle detector, we report the first observation of the decay 
$D_{s1}(2536)^+ \rightarrow D^+ \pi^- K^+$.
The ratio of branching fractions  
$\frac{\mathcal{B}(D_{s1}^+\rightarrow D^+\pi^-K^+)}{\mathcal{B}(D_{s1}^+\rightarrow D^{*+}K^0)}$ is measured to be 
$(3.27\pm 0.18\pm0.37)\%$. 
We also study the angular distributions in the $D_{s1}(2536)^+\rightarrow D^{*+}K^0_S$ decay and 
measure the ratio of D- and S-wave amplitudes.
The S-wave dominates, with a partial width of $\Gamma_S/\Gamma_{\mathrm{total}}=0.72\pm0.05\pm0.01$.
\end{abstract}


\maketitle

\tighten

{\renewcommand{\thefootnote}{\fnsymbol{footnote}}}
\setcounter{footnote}{0}

\section{Introduction}

Two states, $D_{s0}^*(2317)^+$ and $D_{s1}(2460)^+$, have been discovered recently both in continuum $e^+e^-$ annihilation near $\sqrt{s}=10.6$ GeV/$c^2$
 and in $B$ meson decays~\cite{dsj,krok,dspipi}.
Their spin-parities are, respectively, $J^P=0^+$ and $1^+$~\cite{pdg}, and they are presumed to be P-wave excited $c\bar{s}$ states with $j=L+S_{\bar{s}}=1/2$.
Here, $L=1$ is the orbital angular momentum and $S_{\bar{s}}$ is the spin of the light antiquark. However, their masses are unexpectedly low~\cite{dsj_theory}.
This has renewed interest in measurements of P-wave excited charm mesons.

We report the first observation of the decay $\dss\to D^+\pi^-K^+$. (The inclusion of charge-conjugate modes
is implied throughout this paper.) The
$D^+\pi^-$ pair in the final state is the only $D \pi$ combination that cannot come from a $D^*$ resonance: 
$D^{*0}$ mesons can only be produced
virtually here since $M_{D^{*0}} < M_{D^+} + M_{\pi^-}$. 
The new $\dss\to\dppik$ mode reported here 
is only the second observed three-body decay of the $\dss$, 
after $D_s^+ \pi^+ \pi^-$~\cite{dspipi}.

In addition, we have performed an angular analysis of the $\dss\to D^{*+} K^0_S$ mode. 
Heavy Quark Effective Theory (HQET) predicts that for an infinitely heavy $c$-quark this decay of a $J^P=1^+$, $j\!=\!3/2$ 
state should proceed via a pure D-wave~\cite{HQET}.
The corresponding decay of its partner, the $D_{s1}(2460)^+$, which is believed to be a $1^+$, $j\!=\!1/2$ state is energetically forbidden, but if it were
allowed it would proceed via a pure S-wave. 
Since heavy quark symmetry is not exact, the two $1^+$ states can mix with each other,
\begin{equation}
\begin{tabular}{ccc@{\hspace{0mm}}c@{\hspace{1mm}}c@{\hspace{1mm}}c}
$\left.|D_{s1}(2460)^+\right>$&$=$&   & $\cos\theta \left.|^{1/2}E_1\right>$& $+$ &$\sin\theta \left.|^{3/2}E_1\right>$, \\
$\left.|\dss\right>$          &$=$&$-$& $\sin\theta \left.|^{1/2}E_1\right>$& $+$ &$\cos\theta \left.|^{3/2}E_1\right>$, \\
\end{tabular}
\end{equation}
where $\left.|^{1/2}E_1\right>$ and $\left.|^{3/2}E_1\right>$ denote the states with $j\!=\!1/2$ and $j\!=\!3/2$, respectively. 
Note that the coupling via common decay channels
can give a contribution to the mixing that might not be well represented by an orthogonal rotation~\cite{godfrey}. We neglect this possibility in
the expression above. 
If $\theta\ne0$,
an S-wave component can appear in the decay $\dss\to D^* K$. 
Moreover, even if $\theta$ is small,
the S-wave component can give a sizeable contribution to the width because
the D-wave contribution is strongly suppressed by the small energy release in the $\dss\to D^* K$ decay.

The first attempt to decompose S- and D-waves in the analogous decays of the non-strange mesons $D_1(2420)^0\to D^{*+}\pi^-$ and $D_1(2420)^+\to D^{*0}\pi^+$
was reported more than ten years ago by CLEO~\cite{cleo1,cleo2}; currently, no results on the $\dss$ exist. 
Moreover, CLEO's method did not allow the measurement of the ratio of partial widths: it only determined the relation between this ratio and the relative phase 
between the S- and D-wave amplitudes. 
Some information on $\theta$ is obtained from the ratio of electromagnetic decay rates $D_{s1}(2460)^+\to D_s^+\gamma, \ D_s^{*+}\gamma$, since
only the $^1P_1$ state in $D_{s1}(2460)^+$ undergoes an E1 transition to $D_s^+$ and only the $^3P_1$ state to $D_s^{*+}$~\cite{mix_th}.
The bases $\left.|^{j}E_1\right>$ and $\left.|^{2S+1}P_1\right>$ are related by the rotation angle $\theta_0$, where $\tan\theta_0 \!=\! -\!\sqrt{2}$. 
The angle between the bases $\left.|D_{s}^+\right>$ and $\left.|^{2S+1}P_1\right>$ is $\theta+\theta_0$.
The Belle Collaboration studied $D_{s1}(2460)^+\to D_s^+\gamma, \ D_s^{*+}\gamma$ decays using $D_{s1}(2460)^+$ from both $B$ decays~\cite{krok} and from $e^+e^-$ 
annihilation~\cite{dspipi}, and determined the ratio of decay rates to be $0.4\pm0.3$ and $0.28\pm0.17$, respectively. 
According to Ref.~\cite{mix_th}, the average ratio of 
$0.31\pm0.14$ gives the constraint $\tan^2(\theta+\theta_0) = 0.8\pm0.4$.
Detailed knowledge of the mixing is important to test different theoretical models~\cite{godfrey, mix_th, th_models}, 
to fix their parameters and to understand better the nature of $D_{sJ}$ mesons.

Finally, using the $\dss\to\dst\ks$ mode, we have measured the spin alignment of high momentum $\dss$ mesons produced in $e^+e^-$ annihilation. 
Production of excited mesons in the HQET framework is described in Ref.~\cite{falk_peskin}. The fragmentation process is assumed to be so rapid that the
color magnetic forces do not have time to act and thus the spin of the light antiquark in the produced meson 
is uncorrelated with that of the heavy quark.
One consequence of this is that $D^*$ mesons 
with $j = {1\over 2}$ are produced unpolarized. 
This was confirmed with good accuracy by CLEO~\cite{cleo_align} in
$e^+e^-\to c\bar{c}$ events at $\sqrt{s}=10.5$~GeV and was also checked by other experiments~\cite{exp_dst_align}.
Another prediction is that $D^*$ and $D$ mesons are produced according to the number of available helicity states in a 3:1 ratio.
However, experimental data 
from several different production mechanisms
($e^+e^-$, hadroproduction, photoproduction, etc.) give 
an average probability for an S-wave meson to be produced in a vector state of $0.594\pm0.010$~\cite{pv}, 
which is much smaller than the expected value of 0.75.

There are no similar measurements for the P-wave states. Contrary to the $(D,\ D^*)$ case,
HQET predicts that the members of $j\!=\!3/2$ doublet can be produced aligned. The probabilities
for the light degree of freedom to have helicity $-3/2,\ -1/2,\ 1/2,\ 3/2$ are expressed via one parameter $w_{3/2}$ as $\frac{1}{2}w_{3/2},\ 
\frac{1}{2}(1-w_{3/2}),\ \frac{1}{2}(1-w_{3/2}),\ \frac{1}{2}w_{3/2}$, respectively. By adding the $c$-quark spin and resolving the $c\bar{s}$ system into
$1^+$ and $2^+$ states, one can calculate their alignment. 
For $\dss$ the probability of zero helicity is $\rho_{00}=\frac{2}{3}(1-w_{3/2})$. 
A calculation based on perturbative QCD and a nonrelativistic quark model gives
$w_{3/2}=29/114\approx0.254$~\cite{chen_wise} and $\rho_{00}\approx0.497$.
This calculation also predicts the dependence of $w_{3/2}$ on the longitudinal momentum fraction and on the transverse momentum of the meson relative to the heavy quark jet.
The ARGUS analysis of the angular distributions in $D_2^*(2460)\to D\pi$ decay~\cite{argus} gives an upper limit $w_{3/2}<0.24$ at 90\%~CL~\cite{falk_peskin}.
Once $w_{3/2}$ is measured, one can make definite predictions for the angular distributions of the remaining $j\!=\!3/2$ meson decays and check the validity
of HQET.

\section{Selection criteria}

This study is based on a data sample of 462~fb$^{-1}$ collected near the $\Upsilon(4S)$ resonance 
with the Belle detector at the KEKB asymmetric-energy $e^+e^-$ (3.5 on 8~GeV) collider~\cite{KEKB}.
The Belle detector is a large-solid-angle magnetic
spectrometer that
consists of a silicon vertex detector,
a 50-layer central drift chamber, an array of
aerogel threshold Cherenkov counters, 
a barrel-like arrangement of time-of-flight
scintillation counters, and an electromagnetic calorimeter
comprised of CsI(Tl) crystals located inside 
a superconducting solenoid coil that provides a 1.5~T
magnetic field.  An iron flux-return located outside
the coil is instrumented to detect $K_L^0$ mesons and to identify
muons.  The detector
is described in detail elsewhere~\cite{Belle}.
Two inner detector configurations were used. A 2.0 cm beampipe
and a 3-layer silicon vertex detector were used for the first sample
of 155~fb$^{-1}$, while a 1.5 cm beampipe, a 4-layer
silicon detector and a small-cell inner drift chamber were used to record  
the remaining 307~fb$^{-1}$~\cite{Ushiroda}.  

$K^{\pm}$ and $\pi^{\pm}$ candidates are required to originate from the
vicinity of the event-dependent interaction point.
To identify kaons, 
we combine the ionization energy loss ($dE/dx$) 
from the central drift chamber, time of flight and Cherenkov light yield
information for each track to form kaon and pion 
likelihoods ${\mathcal L}_K$ and ${\mathcal L}_{\pi}$, 
respectively,~\cite{pid} and then 
impose the requirement ${\mathcal L}_K / ({\mathcal L}_K + {\mathcal L}_{\pi}) > 0.1$. This requirement has 98\% (97\%) efficiency for a kaon
from $\dss$ (kaon from $D$) and a 12\% (17\%) misidentification probability for a pion with the same momentum.
All unused tracks, whether identified as a kaon or 
not, are treated as pion candidates in what follows.
$K_S^0$ candidates are reconstructed via the $\pi^+\pi^-$ 
decay channel, with a mass within $\pm30\,{\mathrm{MeV}}/c^2$ of the nominal 
$K_S^0$ mass (among other quality requirements).
$D^0$ and $D^+$ mesons are reconstructed using $K^-\pi^+$, $K^0_S \pi^+\pi^-$, $K^-\pi^+\pi^+\pi^-$ and 
$K^0_S \pi^+$, $K^-\pi^+\pi^+$ decay modes, respectively.
All combinations with masses within $\pm20$~MeV/$c^2$ of the nominal $D$ mass are selected (99\% efficiency);
a mass and vertex constrained fit is then applied.

$D^{*+}$ mesons are reconstructed using the $D^0 \pi^+$ mode.
The slow $\pi^+$ momentum resolution is degraded by multiple scattering, but is improved by a track refit procedure 
in which the $\pi^+$ origin point is constrained by the intersection of the $D^0$ momentum and the known $e^+e^-$ interaction region.
The $D^0 \pi^+$ mass is required to be within $\pm 1.5$~MeV/$c^2$ of the $D^{*+}$ nominal value, which corresponds to 
98\% efficiency.  A $D^{*+}$ mass constraint is not imposed.  Instead, we 
characterize the $D_{s1}(2536)^+$ candidate using the mass difference $M_{D^0\pi^+K^0_S} - M_{D^0\pi^+}$, 
where the error in the $D^{*+}$ momentum nearly cancels out.  For the 
$D_{s1}(2536)^+ \to D^+ \pi^- K^+$ decay mode, the track 
refit procedure described above is applied to the pion and kaon momenta, and the 
$D_{s1}(2536)^+$ is characterized by the $D^+ \pi^- K^+$ mass.


It is known that the momentum spectrum of the excited charm resonances from continuum $e^+e^-$ annihilation is hard. 
In addition, due to the strong magnetic field in the Belle detector,
the reconstruction efficiency for slow $\pi^{\pm}$ and $K^+$ mesons rises with $\dss$ momentum.
Therefore, we require $x_P > 0.8$ for the scaled momentum $x_P$,
defined as the ratio $p^*/p^*_{\mathrm{max}}$. 
Here, $p^*$ is the momentum 
of the $\dss$ candidate in the $e^+e^-$ center-of-mass frame, while $p^*_{\mathrm{max}}= \sqrt{E^{* 2}_{\mathrm{beam}}-M^2}$ 
is the momentum in this frame for a candidate carrying all the beam energy.
This selection also removes $\dss$ mesons produced in the decays of $B$ mesons.

In Monte Carlo (MC) simulation, $\dss$ mesons from $e^+e^-$ annihilation, particle decays and the detailed detector response are simulated using the
{\tt PYTHIA}, {\tt EvtGen} and {\tt GEANT} packages~\cite{mcpackages}, respectively.
The $D^0$ and $D^+$ decay modes used in reconstruction are generated with their resonant substructures taken from the Particle Data Group (PDG) compilation~\cite{pdg} but
neglecting any interference effects. 
The $\dss$ momentum spectrum as measured with the $\dss\to\dst\ks$ decay mode is used for MC generation.
As shown below, no clear resonant substructure is visible in the decay $\dss\to D^+\pi^-K^+$. 
Therefore, this mode is simulated as a three-body phase space decay. 
$\dss\to\dst\ks$ decays are generated according to our measured $\dss$ polarization and D/S-wave interference.

\section{$\dss\to D^+\pi^-K^+$ decay}

\begin{figure}[htb]
\includegraphics[width=0.54\textwidth]{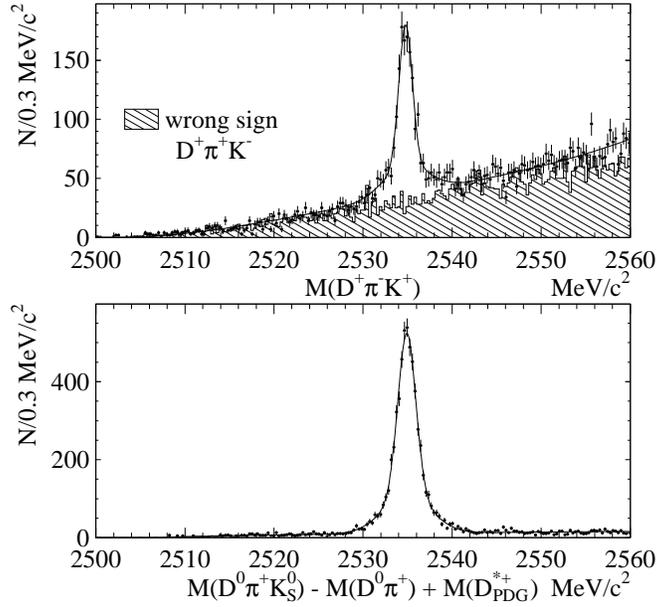}
\caption{$\dss$ mass spectra for $D^+\pi^-K^+$ (top) and $D^{*+}K^0_S$ (bottom) decay modes. 
The hatched histogram in the top plot shows the corresponding spectrum of wrong sign $D^+\pi^+K^-$ combinations.
The fit is described in the text. The fit results are listed in Table~\ref{yields}.}
\label{m}
\end{figure}

The mass $M_{D^+\pi^-K^+}$ (upper plot) and  the mass difference $(M_{D^0\pi^+K^0_S} - M_{D^0\pi^+}) + M^{\rm{PDG}}_{D^{*+}}$ (lower plot) for all
accepted combinations are shown in Fig.~\ref{m}. The PDG superscript denotes the nominal mass value from Ref.~\cite{pdg}.
A clear peak for the new decay channel $\dss\to D^+\pi^-K^+$ is visible.
The mass spectrum of the wrong sign combinations $D^+\pi^+K^-$ shown by the hatched histogram has no enhancement in the $\dss$ region.

To obtain the number of $\dss$ decays, each of the distributions in Fig.~\ref{m} is fit to the sum of two Gaussians with a common mean
(but not necessarily common between the two decay modes).
To ensure that the second Gaussian is always wider than the first one, its width is chosen to be of the form 
$\sigma_2=\sqrt{\sigma_1^2+\Delta\sigma^2}$.
The position of the peak, $\sigma_1$, $\Delta\sigma$, the fraction of events in the first Gaussian and the total number of events in two Gaussians are allowed
to vary in the fit.
The background for the three-body $D^+\pi^-K^+$ (two-body $\dst\ks$) mode is parameterized by a second (first) order 
polynomial multiplied by the threshold 
function $(M-M^{\mathrm{thr}}_{D^+\pi^-K^+})^2$ ($\sqrt{M-M^{\mathrm{thr}}_{D^{*+}K^0_S}}$), where 
$M^{\mathrm{thr}}_f$ is the sum of the nominal masses of final state particles $f$~\cite{pdg}.
Table~\ref{yields} contains the fit results together with the parameters of the Gaussians obtained from MC simulation. 
There is a small fraction of events that contribute two entries to the $\dss$ signal region in the mass plot. 
The excess of such events in comparison with the same number averaged over the left and the right 
sideband is $35-18.5=16.5$ and $203-15.5=187.5$ for the $\dppik$ and $\dst\ks$ modes, respectively. These values are estimates of the contribution of double counted 
signal events. In the branching ratio calculation they are subtracted from the yields given in Table~\ref{yields}.
The signal and the sidebands are defined as 
$|\Delta M_{D_{s1}^+}|<5$ MeV/$c^2$, $10$ MeV/$c^2$ $<|\Delta M_{D_{s1}^+}|<20$ MeV/$c^2$,
respectively, where $\Delta M_{D_{s1}^+}$ is measured relative to the peak position obtained from the fit. 

\begin{table}[htb]
\caption{ Fit results for the $\dss$ spectra in Fig.~\ref{m} and for the corresponding MC simulation spectra: 
number of events in the two Gaussians for data ({\it
 or the efficiency for MC simulation}), fraction of events in the narrow  
 Gaussian, width of the narrow Gaussian, additional width contribution for
 the wide Gaussian, and the mass difference with respect to
 $M_{D_{s1}}^{PDG} = (2535.35 \pm 0.34 \pm 0.5)\,{\mathrm{MeV}}/c^2$.
}
\label{yields}
\begin{tabular}
{@{\hspace{1.5mm}}l@{\hspace{1.5mm}}@{\hspace{1.5mm}}c@{\hspace{1.5mm}}@{\hspace{1.5mm}}c@{\hspace{1.5mm}}@{\hspace{1.5mm}}c@{\hspace{1.5mm}}
@{\hspace{1.5mm}}c@{\hspace{1.5mm}}@{\hspace{1.5mm}}c@{\hspace{1.5mm}}}
\hline
                       & Yield  & Narrow Gaussian & $\sigma_1$ & $\Delta\sigma$ & $M_{D_{s1}}-M_{D_{s1}}^{PDG}$ \\
                       & ({\it Efficiency}) & Fraction & $({\mathrm{MeV}}/c^2)$ &$({\mathrm{MeV}}/c^2)$  & $({\mathrm{MeV}}/c^2)$ \\ \hline
$\dppik$, data  & $\bf 1281\pm66$ & $\bf 0.59\pm0.06$ & $\bf 0.76\pm0.06$ & $\bf 2.4\pm0.4$ & $\bf -0.57\pm0.04$ \\ 
$\dppik$, MC    & $\mathit{0.2699\pm0.0017}$ 
                                  & $0.463\pm0.014$ & $0.94\pm0.02$ & $2.55\pm0.04$ & $-0.031\pm0.010$ \\ \hline
$\dst\ks$, data & $\bf 5673\pm81$ & $\bf 0.63\pm0.03$ & $\bf 1.01\pm0.03$ & $\bf 2.54\pm0.13$ & $\bf -0.43\pm0.02$ \\ 
$\dst\ks$, MC   & $\mathit{0.1273\pm0.0004}$ 
                                  & $0.629\pm0.006$ & $0.946\pm0.008$ & $2.56\pm0.03$ & $-0.034\pm0.005$ \\ \hline
\end{tabular}
\end{table}

The ratio of branching fractions is found to be
\begin{equation}
\frac{\mathcal{B}(\dss\to D^+\pi^-K^+)}{\mathcal{B}(\dss\to D^{*+}K^0)} = 
(3.27\pm 0.18\pm0.37)\%,
\end{equation}
where the first error is statistical and the second is systematic. This 
ratio takes into account the partial branching fractions 
for $D^{(*)}$ mesons to the selected final states and of neutral kaons to 
$\pi^+\pi^-$. 

The systematic error 
receives contribution from the sources listed in Table~\ref{syst}.
A possible difference between the data and MC simulation in evaluation of the
tracking efficiency is estimated using partially reconstructed $D^{*+}$
decays. The reconstruction efficiency errors for slow $K^+$, $\pi^{\pm}$ and $K^0_S$ are added
linearly. The uncertainty in the kaon particle identification is estimated using $D^{*+}$ decays.
Uncertainty in the ratio of $D^+$ and $D^0$ efficiencies is conservatively estimated by a
comparison of different decay modes used in the reconstruction. 
One of the largest contributions to the systematic uncertainty arises due to the
model of the background. It is evaluated by fitting the wrong sign $D^+\pi^+K^-$ subtracted spectrum, which contains almost no background.
For the $D^+\pi^-K^+$ mode, the efficiency is almost independent of the $D^+\pi^-$, $K^+\pi^-$ masses and the angular distribution of decay products.
Therefore, the possible difference between  the simplified phase space MC model and the real $\dss\to\dppik$ decay
results in a small uncertainty in the efficiency determination.
It is estimated by comparing
the yields of events using either an average or differential efficiency in the $\dss$ decay angles and the $D^+\pi^-$ and $K^+\pi^-$ masses.
 The total systematic error is found to be
11.2\% (Table~\ref{syst}). 

\begin{table}[htb]
\caption{Systematic uncertainties for $\mathcal{B}(\dss\to D^+\pi^-K^+)/\mathcal{B}(\dss\to D^{*+}K^0)$.}
\label{syst}
\begin{tabular}
{ @{\hspace{2.4mm}}l@{\hspace{2.4mm}} @{\hspace{2.4mm}}c@{\hspace{2.4mm}} }
\hline\hline
Source & Uncertainty, \% \\ \hline
Reconstruction efficiencies of slow $\pi^{\pm}$, $K^+$ and $\ks$ from $\dst$ and 
$\dss$ & $7.5$ \\
Particle identification of slow $K^+$ from $\dss$ & 1.2 \\
Ratio of $D^+$ and $D^0$ efficiencies& $2.7$ \\
Background model in $M(D^+\pi^-K^+)$ spectrum & 6.5 \\
Efficiency dependence on $D^+\pi^-$, $K^+\pi^-$ masses and & \\
\quad\ angular distribution of decay products in $\dppik$ decay & 1.2 \\
Branching ratio of intermediate resonances\cite{pdg}
                                     & 4.1 \\ \hline
Total                                & 11.2\\ \hline\hline
\end{tabular}
\end{table}

\begin{figure}[htb]
\includegraphics[width=0.54\textwidth]{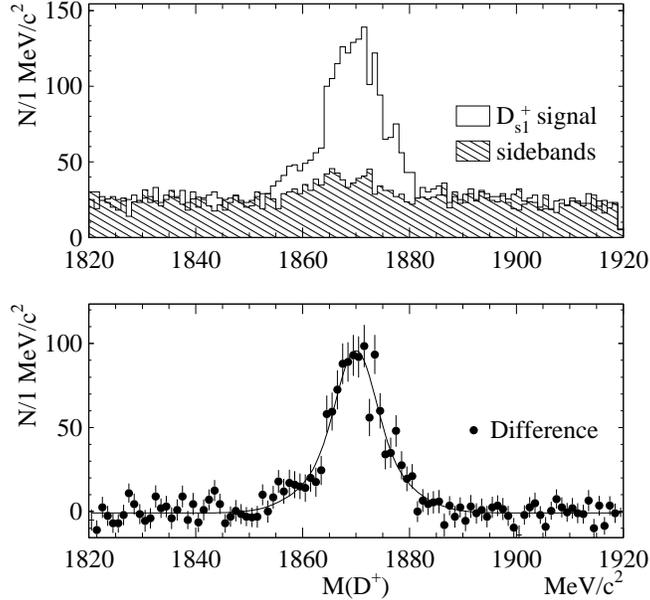}
\caption{$D^+$ mass spectrum for the $\dss$ signal band ($|\Delta M_{D^+\pi^-K^+}|<5$~MeV/$c^2$, open histogram in the top plot) 
and the sidebands ($10<|\Delta M_{D^+\pi^-K^+}|<20$~MeV/$c^2$, normalized to the signal interval, hatched histogram).
$\Delta M_{D^+\pi^-K^+}$ is measured relative to the peak position in the top plot of Fig.~\ref{m}. 
The bottom plot shows the sideband subtracted distribution. The solid curve shows the results of the fit described in the text.}
\label{dp}
\end{figure}

To cross-check the results, the $D^+$ mass spectrum is plotted in Fig.~\ref{dp} for the $\dss$ signal
and sidebands. The latter is normalized to the area of the signal interval.
The sideband subtracted plot shown in the bottom of Fig.~\ref{dp} is fit 
to a double Gaussian as above and a constant background.
The resulting yield $1249 \pm  66$ is consistent with the yield
$1262\pm65$ obtained from the fit of the $\dss$ mass spectrum. 
The constant background level is found to be $-0.9 \pm 0.8$, which is consistent with zero.
The enhancement in the $D^+$ mass region observed 
in the $\dss$ sidebands (top plot of Fig.~\ref{dp}) is due to 
combinations of a real $D^+$ with a random $\pi^- K^+$ pair in the event. 

\begin{figure}[htb]
\includegraphics[width=0.54\textwidth]{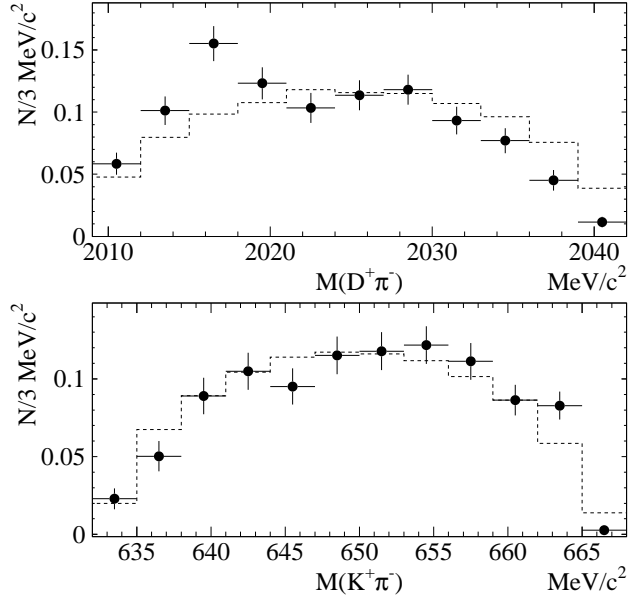}
\caption{Normalized mass spectra of $D^+\pi^-$ (top) and $K^+\pi^-$ (bottom) pairs 
from $\dss\to D^+\pi^-K^+$ decay obtained from fits to the $D^+\pi^-K^+$ mass distributions 
in different $D^+\pi^-$ or $K^+\pi^-$ mass bins.
The dashed histograms show the corresponding MC distributions for $\dss\to\dppik$ decays simulated according to a phase space distribution.}
\label{mdpi_mkpi}
\end{figure}

The $D^+\pi^-$ and $K^+\pi^-$ mass distributions for the 
$\dss\to D^+\pi^-K^+$ decay are shown in Fig.~\ref{mdpi_mkpi}.
The $\dss$ signal yield is obtained from fits to the $D^+\pi^-K^+$ mass 
distribution in bins of $D^+\pi^-$ and $K^+\pi^-$ mass. 
All Gaussian parameters except the total
number of events are fixed in the fit to the values listed in Table~\ref{yields}.
The position of the threshold used for the background description 
depends on the chosen bin.
The areas under the histograms have been normalized to unity. 
The spectra are not efficiency corrected.
The dashed histograms show the corresponding MC spectra for $\dss\to\dppik$ decays simulated according to a phase space distribution.
From this plot, the data points do not appear to be 
entirely consistent with a phase space distribution, but neither do they exhibit any clear dominant resonant substructure.

\section{Angular analysis of $\dss\to D^{*+}K^0_S$ decay}

\begin{figure}[htb]
\includegraphics[width=0.54\textwidth]{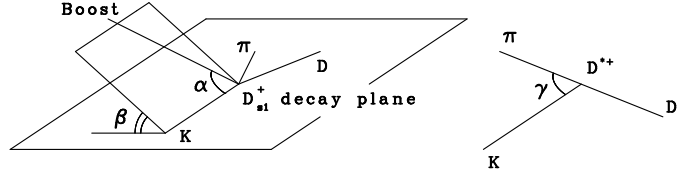}
\caption{Definitions of the angles $\alpha$, $\beta$ and $\gamma$. 
The first two are measured in the $\dss$ rest frame, the third in the $D^{*+}$ frame. 
``Boost'' refers to the direction of the $e^+e^-$ center-of-mass in the $\dss$ rest frame.}
\label{angles_scheme}
\end{figure}

The $\dss\to D^{*+}K^0_S$ decay kinematics can be described by three angles $\alpha$, $\beta$ and $\gamma$ defined as shown in Fig.~\ref{angles_scheme}.
The angles $\alpha$ and $\beta$ are measured in the $D_{s1}^+$
rest frame: $\alpha$ is the angle between the boost direction of the
$e^+e^-$ center-of-mass and the $K_S^0$ momentum, while $\beta$ is the
angle between the plane formed by these two vectors and the $D_{s1}^+$  
decay plane.
The third angle $\gamma$ is defined in the
$D^{*+}$ rest frame between $\pi^+$ and $K^0_S$. 

$\dss$ polarization can be described in terms of its helicity density matrix $\rho_{m_1m_2}$.
The contribution of the element $\rho_{m_1m_2}$ to the decay amplitude is proportional to 
$e^{-i\phi(m_1-m_2)}$, where $\phi$ is the azimuthal rotation angle around 
the $e^+e^-$ boost direction in the $D_{s1}^+$ rest frame.
After integration over $\phi$ the contribution of off-diagonal elements vanishes.
Due to parity conservation, the three diagonal elements can be expressed in terms of the longitudinal polarization $\rho_{00}$, i.e., the
probability that the $D_{s1}(2536)^+$ helicity is zero. 
The other two probabilities are both equal to $\rho_{11}=\rho_{-1-1}=(1-\rho_{00})/2$.
In the helicity formalism, the angular distribution in the decay chain $\dss\to\dst\ks$, $\dst\to D^0\pi^+$ is given by
$$
\frac{d^3N}{d(\!\cos\alpha\!)\,\,d\beta\,\,d(\!\cos\gamma\!)} = \frac{9}{4\pi(1+2R_{\Lambda})}\times
\left(\cos^2\gamma\left[\rho_{00}\cos^2\alpha + \frac{1-\rho_{00}}{2}\sin^2\alpha\right]\right.
$$
$$\left. +
 R_{\Lambda}\sin^2\gamma\left[\frac{1-\rho_{00}}{2}\sin^2\beta + \cos^2\beta(\rho_{00}\sin^2\alpha + \frac{1-\rho_{00}}{2}\cos^2\alpha)\right] \right.
$$
\begin{equation}
+\left.\frac{\sqrt{R_{\Lambda}}\,(1-3\rho_{00})}{4}\sin2\alpha\sin2\gamma\cos\beta\cos\xi\right).
\label{formula}
\end{equation}
The formula depends on three variables: $\rho_{00}$, $R_{\Lambda}$ and $\xi$.
Here $\sqrt{R_{\Lambda}}\,e^{i \xi}=A_{1,0}/A_{0,0} = z$,
where $A_{1,0}$ and $A_{0,0}$ are the helicity amplitudes corresponding to the $D^{*+}$ helicities $\pm1$ and 0, respectively.
They are related to S- and D-wave amplitudes in $\dss$ decay by 
$A_{1,0} = \frac{1}{\sqrt{3}}(S + \frac{1}{\sqrt{2}}D)$, $A_{0,0} = \frac{1}{\sqrt{3}}(S -\sqrt{2}\,D)$. 
Equation~(\ref{formula}) allows one to extract $\rho_{00}$ and $z$ from the $\dss$ angular distributions and to obtain
$D/S = \sqrt{2}\,(z-1)/(1+2z) = \sqrt{\Gamma_D/\Gamma_S}\,e^{i \eta}$, 
where $\Gamma_{D,S}$ are the partial widths of $\dss$ and $\eta$ is the phase between D- and S-amplitudes.

The last interference term in Eq.~(\ref{formula}), with phase $\xi$, vanishes after integration over any angle. In particular, it does not appear in Ref.~\cite{cleo1} in
the formulas for the two- and one-dimensional distributions ${d^2N}/{d(\!\cos\alpha\!)\,d(\!\cos\gamma\!)}$ and ${dN}/{d(\!\cos\gamma\!)}$.
Therefore, in Refs.~\cite{cleo1,cleo2}, only $\sqrt{R_{\Lambda}}=|z|$ is measured for the $D_1(2420)$ meson.
This only constrains the possible ranges of $\Gamma_D/\Gamma_S$ and the phase $\eta$.
To determine them unambiguously, one needs to measure the phase $\xi$
and to fit the whole three-dimensional ${d^3N}/{d(\!\cos\alpha\!)\,\,d\beta\,\,d(\!\cos\gamma\!)}$ distribution.

The probability density function (PDF) for the unbinned maximum likelihood fit has the form
\begin{equation}
{\cal{P}}(\alpha,\,\beta,\,\gamma) = (1-f_b)\cdot\frac{d^3N}{d(\!\cos\alpha\!)\,\,d\beta\,\,d(\!\cos\gamma\!)} 
\cdot \frac{\epsilon(\alpha,\beta,\gamma)}{\left<\epsilon\right>_{\mathrm{avr}}}
 + f_b\cdot{\cal{P}}_{bck}(\alpha,\beta,\gamma).
\label{pdf}
\end{equation}
It includes the efficiency corrections and the contribution of the background.
The background fraction $f_b=528/6169$ is estimated as the ratio of the number of entries in the sidebands 
and in the signal region, respectively.
The signal and the sideband regions, defined as $|\Delta M_{D_{s1}^+}|<7$~MeV/$c^2$ and 10~MeV/$c^2 <|\Delta M_{D_{s1}^+}|<$17~MeV/$c^2$, 
respectively, are wider than in the $\dppik$ case 
since the background is lower.
The PDF ${\cal{P}}_{bck}(\alpha,\beta,\gamma)$, which is normalized to unity, is modelled using the sideband event distribution and the procedure described below. 
$\epsilon(\alpha,\beta,\gamma)$ is the MC-determined efficiency.
The average efficiency $\left<\epsilon\right>_{\mathrm{avr}}$ normalizes to unity the signal part of the PDF.
It is recalculated in every iteration of the fit procedure as 
$\left<\epsilon\right>_{\mathrm{avr}}\equiv \sum_i{\epsilon_i\cdot I_i}
\approx \int\!\!\int\!\!\int \frac{d^3N}{d(\!\cos\alpha\!)\,\,d\beta\,\,d(\!\cos\gamma\!)} \epsilon(\alpha,\beta,\gamma)\,
d(\!\cos\alpha\!)\,\,d\beta\,\,d(\!\cos\gamma\!)$. The sum is taken over 
$10\!\times\!10\!\times\!10$ ``bins'' in a three-dimensional $(\!\cos\alpha\!)\!\times\!\beta\!\times\!(\!\cos\gamma\!)$ space. 
The efficiency map $\epsilon_i$ is determined from
MC simulation, while the integral $I_i$ of ${d^3N}/{d(\!\cos\alpha\!)\,\,d\beta\,\,d(\!\cos\gamma\!)}$ over each bin volume 
is calculated analytically. 

The density ${\cal{P}}_{bck}(\alpha,\,\beta,\,\gamma)$ of sideband events in the
 vicinity of $(\alpha,\beta,\gamma)$ is calculated as follows.  First, the
 three-dimensional $(\!\cos\alpha\!)\!\times\!\beta\!\times\!(\!\cos\gamma\!)$ space is rescaled along each axis to the unit
 cube.  
This ensures that in the case of uniform distributions all three variables have the same ``weight''. 
  Then, for the given point
 $(\alpha,\,\beta,\,\gamma)$, we find the volume $V_{10}$ ($V_{11}$) of the
 smallest cube centered at this point and containing 10 (11) sideband
 events.  
The 10th (11th) event bisects a face of 
$V_{10}$ ($V_{11}$), so the half-weight of event 10 outside 
$V_{10}$ and the half-weight of event 11 inside $V_{11}$ occupy the volume $V_{11}-V_{10}$. Therefore,
we assign 10 full events to the volume $V_{10.5} \equiv {1\over 2}(V_{10} + V_{11})$ and estimate
%
${\cal{P}}_{bck}(\alpha,\,\beta,\,\gamma)=10/V_{10.5}/(4\pi\cdot528)$.  Here, $4\pi$ is the original volume of
 the $(\!\cos\alpha\!)\!\times\!\beta\!\times\!(\!\cos\gamma\!)$ space and 528 is the total number of sideband events.
The resulting ${\cal{P}}_{bck}$ is thus normalized.
To determine a systematic uncertainty due to this procedure, we use 20 or 50 points instead of 10. The changes are found to be
negligible compared to statistical errors (see below).

The advantage of this procedure is that for any one signal event ${\cal{P}}_{bck}$ is always determined from 10 (or 20, or 50) sideband events. 
Therefore the sideband fluctuations are much smaller than fluctuations of one signal event. 
This ensures the necessary degree of ``smoothness'' of the ${\cal{P}}_{bck}$ distribution. 
On the other hand, the typical volume $V_{10.5}$ is about $10/528\approx0.02$ of the whole $(\!\cos\alpha\!)\!\times\!\beta\!\times\!(\!\cos\gamma\!)$         
space volume. ${\cal{P}}_{bck}$ thus reproduces the background behaviour at this level of granularity. 

A similar method is used to construct the efficiency function $\epsilon(\alpha,\,\beta,\,\gamma)$. 
Due to the much larger MC sample, instead of $V_{10.5}$ we use the volume with 100 MC reconstructed events $V_{100.5}$.
We then determine the number of MC events generated there, $N^{100}_{\mathrm{gen}}$, and calculate the
 efficiency as $\epsilon(\alpha,\,\beta,\,\gamma) = 100/N^{100}_{\mathrm{gen}}$.
As in the previous case, usage of the 50 or 200 closest events instead of 100 reproduces the same results within the statistical errors, 
and is used to determine the systematic uncertainty of this method due to the efficiency.
%

The three-dimensional fit of all $\dss$ signal entries to ${\cal{P}}(\alpha,\beta,\gamma)$ gives
\begin{equation}
z=A_{1,0}/A_{0,0} = \sqrt{R_{\Lambda}}e^{i \xi}=\sqrt{3.6 \pm 0.3 \pm 0.1}\exp{(\pm i\cdot(1.27 \pm 0.15 \pm 0.05)}).
\end{equation}
Note that the angular distributions are sensitive only to $\cos\xi$, not to $\xi$ itself. Therefore $\xi$ has a $\pm\xi+2\pi n$ ambiguity, 
and $A_{1,0}/A_{0,0}$ is determined up to complex conjugation. 
The average $\dss$ longitudinal polarization in the region $x_P\!>\!0.8$ is measured to be $\rho_{00}=0.490 \pm 0.012 \pm 0.004$. 

Systematic uncertainties are calculated as a sum in quadrature of the contributions listed in Table~\ref{ang_err}.
MC simulation shows that the detector resolution in $\alpha$, $\beta$, $\gamma$ not only increases
the final errors but also effectively decreases the parameter $R_{\Lambda}$ by 0.13. 
The corresponding correction has already been applied to the above result.
The systematic uncertainties in modelling ${\cal{P}}_{bck}$ and $\epsilon$ are estimated by 
varying the number of closest points as explained above, and by using different sidebands: $7~\mev<|\Delta M_{\dss}|<10~\mev$ plus                       
$17~\mev<|\Delta M_{\dss}|<21~\mev$ (instead of  $10~\mev<|\Delta M_{\dss}|<17~\mev$). 
The errors due to statistical fluctuation of the MC sample, 
which is 15.4 times larger than data and has almost the same $R_{\Lambda},\ \xi$ and $\rho_{00}$,
are calculated as $\sqrt{1/15.4}=0.25$ of the statistical errors.

\begin{table}[htb]
\caption{Systematic uncertainties for $R_{\Lambda}$, $\xi$ and $\rho_{00}$.
}
\label{ang_err}
\begin{tabular}
{ @{\hspace{2.4mm}}l@{\hspace{2.4mm}} @{\hspace{2.4mm}}c@{\hspace{2.4mm}}  @{\hspace{2.4mm}}c@{\hspace{2.4mm}} 
                                      @{\hspace{2.4mm}}c@{\hspace{2.4mm}}  @{\hspace{2.4mm}}c@{\hspace{2.4mm}}}
\hline\hline
Source & $R_{\Lambda}$ & $\xi$ & $\rho_{00}$ \\ \hline
Angular resolution & 0.05 & 0.02 & 0.001 \\
Modeling of ${\cal{P}}_{bck}$ and efficiency & 0.04 & 0.00 & 0.001 \\ 
Different sidebands & 0.03 & 0.02 & 0.002 \\
MC statistics & 0.07 & 0.04 & 0.003 \\ \hline
Total        & 0.10 & 0.05 & 0.004\\ \hline\hline
\end{tabular}
\end{table}

The ratio of D- and S-wave amplitudes is found to be
$D/S = (0.63\pm0.07\pm0.02)\cdot \exp{(\pm i\cdot(0.76\pm0.03\pm0.01))}$.
The relative phase is close to $\pi/4$,
$(43.8\pm1.7\pm0.6)^\circ$.
One can see that, contrary to the HQET prediction, the S-wave dominates. 
Its contribution to the total width is $1/(1+|D/S|^2)=\Gamma_S/\Gamma_{\mathrm{total}}=0.72\pm0.05\pm0.01$.

\begin{figure}[htb]
\includegraphics[width=0.54\textwidth]{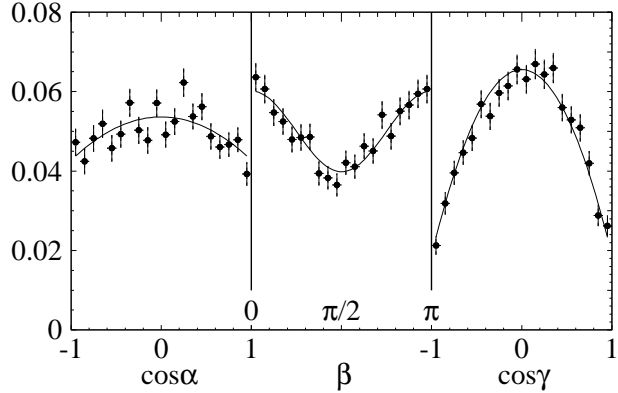}
\caption{Background subtracted, efficiency corrected and normalized one-dimensional $\cos\alpha$, $\beta$, $\cos\gamma$ angular distributions. 
Projected results of the three-dimensional fit are shown by the solid curves.}
\label{abg}
\end{figure}

The background-subtracted efficiency corrected and normalized one-dimensional projections of 
${d^3N}/{d(\!\cos\alpha\!)\,\,d\beta\,\,d(\!\cos\gamma\!)}$ distribution, together with the fit results, are shown in Fig.~\ref{abg}.
The $\chi^2$ difference between the points and the projected fit results corresponds to a goodness-of-fit probability of about 60\%.
As mentioned earlier, one-dimensional projections are not sensitive to the phase $\xi$. They are described instead 
by the following formulas derived from Eq.~(\ref{formula}):
\begin{equation}
\frac{dN}{d\cos\alpha} = \frac{3}{4(1+2R_{\Lambda})} \{\left[(1+R_{\Lambda})+(R_{\Lambda}-1)\rho_{00}\right] + \cos^2\alpha(R_{\Lambda}-1)(1-3\rho_{00})\},
\end{equation}
\begin{equation}
\frac{dN}{d\beta} = \frac{1}{\pi(1+2R_{\Lambda})}\{\left[1+3R_{\Lambda}(1-\rho_{00})\right] + 2R_{\Lambda}(3\rho_{00}-1)\cos^2\beta\},
\end{equation}
\begin{equation}
\frac{dN}{d\cos\gamma} = \frac{3}{2(1+2R_{\Lambda})}\{\cos^2\gamma+R_{\Lambda}\sin^2\gamma\}.
\end{equation}

In spite of the complexity of Eq.~(\ref{formula}), it depends only quadratically on $\sqrt{R_{\Lambda}}$
and only linearly on $\cos\xi$ and $\rho_{00}$. 
The efficiency entering the PDF in Eq.~(\ref{pdf}) is almost constant in all three projections.
When it is set to a constant value in the fit, the results change by less than 1/3 of the statistical error.
The background fraction $f_b=9\%$ is small. Therefore one does not expect any significant biases of the fit results.

To quantify this statement, 1000 samples of events are generated according to PDF Eq.~(\ref{pdf}). 
Each sample contains the same number of events as observed in data.
The parameters $R_{\Lambda},\ \xi$ and $\rho_{00}$ are set
to the values determined from data. A three-dimensional fit is performed for each sample; it is verified that the fit results are not systematically biased and
the errors are estimated correctly. The value of the overall likelihood function is measured 
to be worse than the one observed in data in 33\% of cases. 
                
\begin{figure}[htb]
\includegraphics[width=0.54\textwidth]{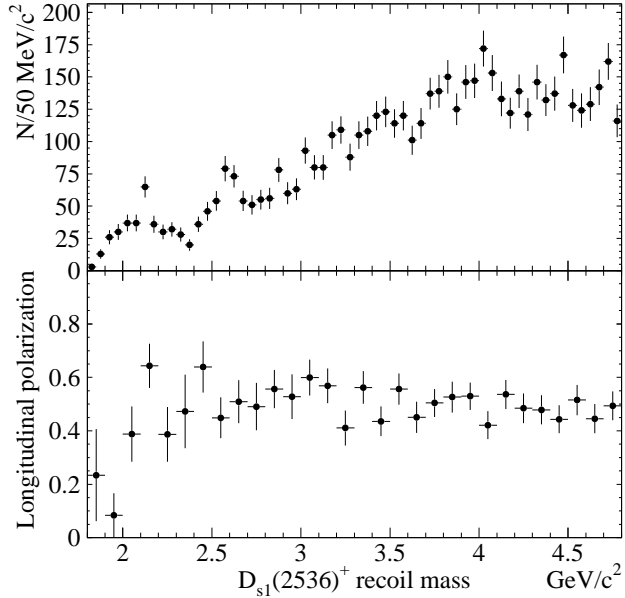}
\caption{$\dss$ recoil mass spectrum (top), probability $\rho_{00}$
that $\dss$ helicity is zero (bottom).
The decrease at low values in the bottom plot can be attributed to the contribution of two-body $e^+e^-\to\dss X$ reactions.}
\label{rm}
\end{figure}

As a final check, the fit to data is repeated in different bins of the mass recoiling against the $\dss$, defined as 
$\sqrt{(2E_{\mathrm{beam}}^*-E_{D_{s1}^+}^*)^2-(p_{D_{s1}^+}^*)^2}$, where all quantities are measured in the $e^+e^-$ center-of-mass frame.
The parameters $R_{\Lambda}$ and $\xi$ are found to be independent of $\dss$ momentum or recoil mass within statistical errors.
The recoil mass spectrum is shown in the top half of Fig.~\ref{rm}.
The resolution is about 70~MeV/$c^2$ at 2~GeV/$c^2$ and is approximately inversely proportional to the recoil mass.
There is an indication of two-body contributions from $e^+e^-\to\dss X$ where $X=D_s^+$, $D_s^{*+}$ and higher $D_s^{**+}$ resonances.
This agrees with the $\dss$ polarization spectrum, shown in the bottom half of Fig.~\ref{rm},
which also exhibits some structure at low recoil masses.
This spectrum 
is obtained 
when $R_{\Lambda}$ and $\xi$ are fixed to their values determined from the overall fit.
At $D_s^+$ mass, as one expects, the longitudinal polarization is low. It then rises rapidly at the $D_s^{*+}$ mass 
and eventually reaches a plateau at $\rho_{00}\approx0.5$.

\section{Conclusions}
In conclusion, a new decay channel $\dss\to D^+\pi^-K^+$ is observed. The $D^+\pi^-$ pair is the only $D\pi$ combination that cannot come
from a $D^*$ resonance. It can be produced only through the virtual $D^{*0}$, broad $D^{*0}_0$ or $D^{*}_2(2460)^0$  resonances. 
In addition, the $D^+\pi^-K^+$ final state can be formed by two-body decays to a $D^+$ and a virtual $K^{*0}$ or higher $K^*$ resonance. 
No clear resonant substructure is found in the $D^+\pi^-K^+$ system. 
The ratio of branching fractions  
${\mathcal{B}(D_{s1}^+\to D^+\pi^-K^+)}/{\mathcal{B}(D_{s1}^+\to D^{*+}K^0)}$ is measured to be 
$(3.27\pm 0.18\pm0.37)\%$. 

An angular analysis of the decay $\dss\to D^{*+}K^0_S$ is also performed. 
Since the $c$-quark is not infinitely heavy, HQET is violated and the $\dss$ can contain an admixture of another $J^P=1^+$ state with $j\!=\!1/2$ and can decay
in an S-wave.
The energy release in this reaction is small. 
Therefore the D-wave is suppressed by the barrier factor $(q/q_0)^5$, where $q$ is the relative momentum of $\dss$ decay products in the $\dss$ rest frame, and $q_0$ is
a momentum scale characteristic of the decay. The S-wave contribution to the total width is proportional to $q/q_0$ and 
can be sizeable even if the mixing is small.
Using an unbinned maximum likelihood fit to the three angles in the $\dss\to\dst\ks$, $\dst\to D^0\pi^+$ decay chain, 
we measure (up to a complex conjugation) the ratio of S- and D-wave amplitudes:
$D/S = (0.63\pm0.07\pm0.02)\cdot \exp{(\pm \,i\cdot(0.76\pm0.03\pm0.01))}$. 
The S-wave dominates, and its contribution to the total width is $\Gamma_S/\Gamma_{\mathrm{total}}=0.72\pm0.05\pm0.01$. This result allows to calculate the mixing
angle in the theoretical models with a known value of parameter $q_0$~\cite{godfrey, th_models}.

The spin of high momentum $\dss$ mesons produced in $e^+e^-$ annihilation prefers to align transversely to the momentum.
The probability that a $\dss$ with $x_P\!>\!0.8$ has zero helicity is found to be $\rho_{00}=0.490 \pm 0.012 \pm 0.004$. 
Assuming the HQET relation $\rho_{00}=\frac{2}{3}(1-w_{3/2})$,~\cite{falk_peskin} this implies 
a value of the Falk--Peskin parameter, $w_{3/2}=0.266 \pm 0.018 \pm 0.006$, in this momentum region.
This value is close to the prediction of Ref.~\cite{chen_wise}, $w_{3/2}\approx0.254$, obtained for the entire momentum region, 
although the applicability of the perturbative QCD fragmentation model for $D_s^{**+}$ mesons is questionable.

\section{Acknowledgments}
We thank the KEKB group for the excellent operation of the
accelerator, the KEK cryogenics group for the efficient
operation of the solenoid, and the KEK computer group and
the National Institute of Informatics for valuable computing
and Super-SINET network support. We acknowledge support from
the Ministry of Education, Culture, Sports, Science, and
Technology of Japan and the Japan Society for the Promotion
of Science; the Australian Research Council and the
Australian Department of Education, Science and Training;
the National Science Foundation of China and the Knowledge
Innovation Program of the Chinese Academy of Sciences under
contract No.~10575109 and IHEP-U-503; the Department of
Science and Technology of India; 
the BK21 program of the Ministry of Education of Korea, 
the CHEP SRC program and Basic Research program 
(grant No.~R01-2005-000-10089-0) of the Korea Science and
Engineering Foundation, and the Pure Basic Research Group 
program of the Korea Research Foundation; 
the Polish State Committee for Scientific Research; 
the Ministry of Education and Science of the Russian
Federation and the Russian Federal Agency for Atomic Energy;
the Slovenian Research Agency;  the Swiss
National Science Foundation; the National Science Council
and the Ministry of Education of Taiwan; and the U.S.\
Department of Energy.

\end{document}